\documentclass[aps,prd,amssymb,eqsecnum,nofootinbib,a4paper]{revtex4}   
\usepackage{helvet,amsmath,amssymb,amsfonts,bm,yhmath,bbm} 
\newif\ifusesec
\usesectrue 
\usepackage{geometry}                
\usepackage{graphics}    
\usepackage{epstopdf}
\usepackage{helvet,amsmath,amssymb,amsfonts,amsthm,bm} 
\usepackage[latin1]{inputenc} 
\usepackage[T1]{fontenc} 
\usepackage[english]{babel}
\usepackage{graphicx,epsfig,stackrel} 
\usepackage[all]{xy}
\usepackage{hyperref}
\usepackage{tablefootnote}
\usepackage{accents}

\newcommand{\ft}[2]{{\textstyle\frac{#1}{#2}}}
 \usepackage{ulem}
\begin{document}
\title{   {Comment on ``Spherically symmetric perturbations of a Schwarzschild
   black hole in torsion bigravity''} }
\author{Philippe Spindel}
\email{philippe.spindel@umons.ac.be}
\affiliation{Physique de l'Univers, Champs et Gravitation, Universit\'e de Mons, Facult\'e des Sciences,20, Place du Parc, B-7000 Mons, Belgium\\
Service de Physique th\'eorique, CP 225, ULB, \\Bld du Triomphe, 1050 Brussels, Belgium}
\date{\today}

\date{\today}

\begin{abstract}
The goal of this short note is to provide a simpler derivation of the effective potential surrounding a Schwarzschild black hole for spherically symmetric perturbations in the framework of torsion bigravity than the one presented in Ref.\cite{VN}. We also discuss the unicity of the reduction process that leads to the potential.
\end{abstract}


\maketitle

\section{Introduction}
In a recent work \cite{VN}, a Zerilli-like equation driving perturbations around  a Schwarzschild black hole was established and discussed in the framework of torsion bigravity theory. In our opinion the choice of the variables adopted   makes the obtention of the perturbation equation a real ``{\it tour de force}''. Hereafter we  indicate a more straightforward way that leads to the same result.
\section{Reduction}
Instead of using Cartan formalism we  adopt the metric and connection approach  based on the introduction of Lagrange multipliers as described in  Ref. \cite{PS}. The corresponding field equations are given by Eqs (2.17)--(2.19),  specialised to the Lagrangian Eq.~(2.8) with $L_F$ given by Eq.~(3.1).  The various coupling constants expressed in terms of  the parametrisation  used in Ref. \cite{VN} are :
\begin{align}
&c_R=\frac{\lambda}{(1+\eta)}\ ,\  c_F=\frac{\eta\,\lambda}{(1+\eta)}\ ,\  d_3=2\,\frac{\eta\,\lambda}{\kappa^2}\ ,\  f_1=\frac{\eta\,\lambda}{\kappa^2}+c_{34}\ ,\  f_2=-\frac 53\,\frac{\eta\,\lambda}{\kappa^2}-c_{34}\ ,\\
&d_1=d_2=0\ ,\   \Lambda=0\  .
\end{align}
The geometrical entities are :
\begin{itemize}
\item{The metric :}
\begin{align}
ds^2=-e^{ 2\,\Phi[t,r]}\,dt^2+e^{2\,\Lambda[t,r]}\,dr^2+r^2\,(d\theta^2+\sin^2[\theta]\,d\varphi^2)
\end{align}
\item The non-zero even-parity torsion\footnote{In this note, torsion is defined according to Cartan structure equation (for the notations, see Ref.\cite{PS}, Section {\bf II}.C): $\mathbf T^{\hat a}=\ft 12\,T^{\hat a}_{\phantom{a}\hat b\hat c}\,\underline e^{\hat b}\wedge \underline e^{\hat c}=d\underline e^{\hat a}+\underline A^{\hat a}_{\phantom{a}\hat b}\wedge \underline e^{\hat b}$; in Ref.\cite{VN} the opposite convention is used.}  components :
\begin{align}
&T_{rrt}=e^{2\,\Lambda[t,r]}(\dot \Lambda[t,r]-e^{\Phi[t,r]}\,X[t,r]) \qquad , \\
&T_{\theta r\theta}=(r+r^2\,e^{\Lambda[t,r]}\,W[t,r])\qquad, \qquad T_{\varphi r \varphi}=\sin^2[\theta]\,T_{\theta r \theta} \qquad , \\
&  {T_{\theta \theta t}}=r^2\,e^{\Phi[t,r]}\,Y[t,r]\qquad\qquad \quad, \qquad   {T_{\varphi   \varphi t}}=\sin^2[\theta]\,T_{\theta \theta t} \qquad , \\
&T_{t r t}=e^{2\,\Phi[t,r]}(e^{\Lambda[t,r]}\,V[t,r]-\Phi'[t,r])\qquad ,
\end{align}
and those related by symmetry :$T_{\alpha\mu\nu}=-T_{\alpha\nu\mu}$. As shown in Ref. \cite{VN}, odd-parity components can be consistently set equal zero from the beginning of the discussion.
\end{itemize}
The links between the   field equations written directly in terms of the natural components of the metric and the connection and those written in terms of frame components are (using the numbering of Appendix A of Ref.\cite{VN} and the notations of Section {\bf II} of Ref.\cite{PS})   :
$\text{Eq. (A4)}\sim \mathcal E^t_t$, $\text{Eq. (A5)}\sim \mathcal E^r_t$, $\text{Eq. (A6)}\sim \mathcal E^\theta_\theta$, the sum : $\text{Eq. (A7)+Eq. (A8)}\sim \mathcal E^t_r$, $\text{Eq. (A9)}\sim \mathcal S^{rtt}$, $\text{Eq. (A10)}\sim \mathcal S^{rtr}$, $\text{Eq. (A11)}\sim \mathcal S^{\theta t\theta}$, $\text{Eq. (A12)}\sim \mathcal S^{r\theta\theta}$. The difference : $\text{Eq. (A7)-Eq. (A8)}$ is a consequence of the previous equations (See Eq. (2.42) of Ref.\cite{PS}). Thus we are confronted {\it a priori} to a system of only 8 even-parity equations.

To pursue the analysis we introduce metric and torsion perturbations around the Schwarzschild background as follows :
\begin{align}
&\Lambda[t,r]=-\ft 12\log(1-\frac{2\,m}r)+\epsilon\int_{-\infty}^{+\infty} e^{-i\,\omega\,t}a[\omega,r]\,d\omega\label{Lampert} \qquad , \\
&\Phi[t,r]=\ft 12\log(1-\frac{2\,m}r)+\epsilon\int_{-\infty}^{+\infty} e^{-i\,\omega\,t}b[\omega,r]\,d\omega \qquad , \\
&V[t,r]=e^{-\Lambda[t,r]}\,\Phi'[t,r]+\epsilon\int_{-\infty}^{+\infty} e^{-i\,\omega\,t}v[\omega,r]\,d\omega \qquad , \\
&W[t,r]=-  {\frac{e^{-\Lambda[t,r]}}r}+\epsilon\int_{-\infty}^{+\infty} e^{-i\,\omega\,t}w[\omega,r]\,d\omega \qquad , \\
&X[t,r]=e^{-\Phi[t,r]}\,\dot\Lambda[t,r]+\epsilon\int_{-\infty}^{+\infty} e^{-i\,\omega\,t}x[\omega,r]\,d\omega \qquad , \\
&Y[t,r]=\ +\epsilon\int_{-\infty}^{+\infty} e^{-i\,\omega\,t}y[\omega,r]\,d\omega\label{Ypert}\qquad .
\end{align}
These perturbations variables are related to those introduced in Ref.\cite{VN} by : 
\begin{align}
&\Lambda_o[\omega,r]=a[\omega,r] \qquad , \\
&\phi_o[\omega,r]=b[\omega,r] \qquad , \\
&V_o[\omega,r]=v[\omega,r]+\sqrt{1-\frac {2\,m}r}\big(b'[\omega,r]-\frac m {r\,(r-2\,m)}\,a[\omega,r]\big) \qquad , \\
&W_o[\omega,r]=w[\omega,r]-\frac 1r\,\sqrt{1-\frac {2\,m}r}\,a[\omega,r] \qquad , \\
&X_o[\omega,r]=x[\omega,r]-i\,\omega\,\frac 1{\sqrt{1-\frac {2\,m}r}}\,a[\omega,r] \qquad , \\
&Y_o[\omega,r]=y[\omega,r]\qquad .
\end{align}
Plugging the expressions \ref{Lampert}--\ref{Ypert} into the field equations and expanding them to first order in $\epsilon$ provides the perturbation equations. They consist of 8 equations (equivalent to those provided in the {\it Supplemental Material} linked to Ref. \cite{VN}). Two of them involve the third derivative (with respect to $r$) of $b[\omega,r]$ and the second derivatives of all the 6 perturbations variables. The redundancy of this system allows one, by taking linear combinations of the various equations and their derivatives, to reduce the system to :
\begin{align}
&v[\omega,r]=2\frac{\kappa^2r^3+m\,(1+\eta)}{\kappa\,r^3-2\,m\,(1+\eta)}w[\omega,r]\qquad,\\
&x[\omega,r]=-2\frac{\kappa^2r^3+m\,(1+\eta)}{\kappa\,r^3-2\,m\,(1+\eta)}y[\omega,r]\qquad,\\
&a[\omega,r]=\sqrt{1-\frac{2\,m}r}\frac \eta{1+\eta}\,\frac{\kappa^2\,r^3-2\,m\,(1+\eta)}{(r-2\,m)}{(\kappa^2\,r^3-2\,m\,\eta)}\Big (r\,w[\omega,r]+i\,\frac m{\omega\,r}\,y[\omega,r]\Big)\qquad,\\
&b'[\omega,r]= -\frac1{\sqrt{1-\frac{2\,m}r}}\Big(\frac{\eta r \big(\kappa^2 r^3-2(1+\eta)m\big)\big(\kappa^2 r^3 (2 r-3 m)+2 m\eta(  r-3 m)\big) }{{ (r-2m)(1+\eta)(\kappa^2\,r^3-2\eta \,m)^2}}w[\omega,r]\nonumber\\
&\phantom{b'[\omega,r]=}+\frac{i \eta(\kappa^2 r^3-2m(1+\eta))( \omega^2\, r^3(\kappa^2 r^3-2\, m\,\eta)
+2\,m(\kappa^2 r^2 (2\, r-3\,m)-  \eta\, m)}{ (r-2m)(1+\eta)(\kappa^2\,r^3-2\eta \,m)^2 }y[\omega,r]\Big)\qquad,\\
&w'[\omega,r]=\frac 1{ r}\Big( \frac {r-3\,m}{r-2\,m}+6\,m\,\frac{(1+3\,\eta)\,\kappa^6\,r^9-10\,\eta\,(1+\eta)\,\kappa^4\,m\,r^6+4\,\eta\,(1+\eta)^2\kappa^2\,m^2\,r^3+8\,\eta^2(1+\eta)^2\,m^3}{(\kappa^4\,r^6-4\,\eta\,(1+\eta)\,m^2)(\kappa^2\,r^3-2\,(1+\eta)\,m)(\kappa^2\,r^3-2\eta\,m)}\Big)\,w[\omega,r]\nonumber\\
&\phantom{w'[\omega,r]=}-i\Big(\frac{\kappa^2(\kappa^2\,r^3-2\,(1+\eta)\,m)^2(\kappa^2\,r^3+4\,\eta\,m)}{(\kappa^4\,r^6-4\,\eta\,(1+\eta)\,m^2)(\kappa^2\,r^3-2\,\eta\,m)}-\frac r{r-2\,m}\,\omega^2\Big)q[\omega,r] \label{Eqdw}\qquad,\\
&  {q}'[\omega,r] =i\frac r{r-2\,m}\,w[\omega,r]-\frac{\kappa^2\,r^3\,(3\,r-5\,m)-2\,(1+\eta)\,m^2}{r\,(r-2\,m)(\kappa^2\,r^3-2\,(1+\eta)\,m)} q[\omega,r] \label{Eqdk}\qquad,
\end{align}
where 
\begin{equation}q[\omega,r]:=\frac{y[\omega,r]}\omega\qquad. \end{equation}
\section{Zerilli's construction}
Zerilli has presented \cite{FZ}, in the framework of the analysis of gravitational perturbations around a black hole,   the result of the reduction of a system of first order differential equation system :
\begin{align}
&z_1'[\omega,r]=\alpha[\omega,r]\,z_1[\omega,r]+\beta[\omega,r]\,z_2[\omega,r]\label{Eqsyst1}\\
&z_2'[\omega,r]=\gamma[\omega,r]\,z_1[\omega,r]+\delta[\omega,r]\,z_2[\omega,r]\label{Eqsyst2}
\end{align}
 to a second order Schr\"odinger-type equation :
 \begin{align}
& n[ r]\, \phi'[\omega,r]=\psi[\omega,r]\qquad ,\label{Sch1}\\
&n[ r]\, \psi'[\omega,r]=(V[ r]-\omega^2)\,\phi[\omega,r]\qquad ,\label{Sch2}
\end{align}
 under the assumptions that : 
\begin{align}
& \alpha[\omega,r]=\alpha_0[r]+\alpha_2[r]\,\omega^2\quad,\quad & \beta[\omega,r]=\beta_0[r]+\beta_2[r]\,\omega^2\quad,\label{Eqab}\\
& \gamma[\omega,r]=\gamma_0[r]+\gamma_2[r]\,\omega^2\quad,\quad & \delta[\omega,r]=\delta_0[r]+\delta_2[r]\,\omega^2\quad ,\label{Eqcd}
\end{align}
by defining\footnote{In Refs. \cite{VN,FZ} the transformation considered is  the inverse of the one introduced here : \begin{align}
&  z_1[\omega,r]=f[r]\,\phi[\omega,r]+g[r]\,\psi[\omega,r]\qquad ,\nonumber\\
& z_2\omega,r]=h[ r]\,\phi[\omega,r]+k[r],\psi[\omega,r]\qquad ,\nonumber
\end{align} 
} : 
\begin{align}
&  \phi[\omega,r]=A[r]\,z_1[\omega,r]+B[r]\,z_2[\omega,r]\qquad ,\label{phiz12}\\
& \psi[\omega,r]=C[ r]\,z_1[\omega,r]+D[ r]\,z_2[\omega,r]\qquad . \label{psiz12}
\end{align}
with
\begin{align}
A[r]\,D[r]-B[r]\,C[r]\neq 0\label{detcond}\qquad.
\end{align}
As we don't completely agree on one point exposed in Ref.\cite{FZ}, let us reconsider hereafter the reduction process. \newline
  {Inserting the expressions [\ref{phiz12}, \ref{psiz12}] of $\phi[\omega,r]$ and $\psi[\omega,r]$ into Eqs [\ref{Sch1}, \ref{Sch2}] , taking into account Eqs [\ref{Eqsyst1}, \ref{Eqsyst2}] and the arbitrariness of the values on a point  of the functions $z_1[\omega,r]$ and $z_2[\omega,r]$ (expressing the Cauchy problem of the differential system) we obtain four constraints :
\begin{align}
\text{ {\it $\bullet$ from Eq. }[\ref{Sch1}] :}&\nonumber \\
 &n[ r]\,(A'[ r]+\alpha[\omega, r]\,A[ r]+\gamma[\omega, r] \,B[ r])=  C[ r] \qquad , \label{EqC}\\
 &n[ r]\,(B'[ r]+\delta[\omega, r]\,B[ r]+\beta[\omega, r] \,A[ r])= D[r] \qquad ,\label{EqD}\\
 \text{{\it $\bullet$  from Eq. }[\ref{Sch2}] :}&\nonumber \\
 &n[ r]\,(C'[ r]+\alpha[\omega, r]\,C[ r]+\gamma[\omega, r]\, D[ r])= (V[r] -\omega^2)\,A[r]\qquad ,\label{EqA}\\
 &n[ r]\,(D'[ r]+\beta[\omega, r]\,C[ r]+\delta[\omega, r]\, D[ r])= (V[r] -\omega^2)\,B[r]\label{EqB}\qquad .
\end{align}
They can be rewritten as :}
\begin{align}
&(A'[r]+\alpha[\omega,r]\,A[r]+\gamma[\omega,r]\,B[r])D[r]-(B'[r]+\delta[\omega,r]\,B[r]+\beta[\omega,r]\,A[r])C[r]=0\label{EqAB}\quad , \\
&(C'[r]+\alpha[\omega,r]\,C[r]+\gamma[\omega,r]\,D[r])B[r]-(D'[r]+\delta[\omega,r]\,D[r]+\beta[\omega,r]\,C[r])A[r]=0\label{EqCD}\quad , \\
&n[r]\,\frac{ (B'[r]+\delta[\omega,r]\,B[r]+\beta[\omega,r]\,A[r])A[r]-(A'[r]+\alpha[\omega,r]\,A[r]+\gamma[\omega,r]\,B[r])B[r]}{A[r]\,D[r]-B[r]\,  C[r]}=1\label{EqUn}\quad , \\
&n[r]\,\frac{(C'[r]+\alpha[\omega,r]\,C[r]+\gamma[\omega,r]\,D[r])D[r]-(D'[r]+\delta[\omega,r]\,D[r]+\beta[\omega,r]\,C[r])C[r]}{A[r]\,D[r]-B[r]\,C[r]}=V[r]-\omega^2\label{EqVw}\quad .
\end{align}

Combining Eqs [\ref{EqAB}] and [\ref{EqCD}] we obtain :
\begin{align}
\frac{(A[ r]\,D[ r]-B[ r]\,C[ r])'}{(A[ r]\,D[ r]-B[ r]\,C[r])}=-(\alpha[\omega,r]+\delta[\omega,r])\label{dlogADBC}
\end{align}
whose independence with respect to $\omega$ requires that :
\begin{align}
&\alpha[\omega,r]+\delta[\omega,r]= \alpha_0[ r]+\delta_0[ r]
\end{align}
 {\it i.e.} 
 \begin{align}
& \alpha_2[ r]=-\delta_2[ r]\label{conda2d2}
\end{align}
  Thus   { the integration of  Eq.  [\ref{dlogADBC}] leads  to a first relation :}
\begin{align}
 (A[ r]\,D[ r]-B[ r]\,C[ r]) =   e^{-\int  (\alpha_0[ r]+\delta_0[ r])\,dr}\qquad .\label{ADBC}
 \end{align}
   {We obtain from Eqs [\ref{EqC}, \ref{EqD}] that the independence of the functions }$C[ r]$ and $D[ r]$ with respect to $\omega$ requires :
  \begin{align}
& \alpha_2[r] \,A[ r]+ {\gamma_2[ r]} \,B[ r]=0\qquad,\label{aAgB}\\
& \beta_2[r] \,A[ r]+ {\delta_2[ r]} \,B[ r]=0\label{bAdB}\\
 \end{align}
 which implies the compatibility condition :
 \begin{align}
& \alpha_2[r] \,{\delta_2[ r]}- \beta_2[r] \, {\gamma_2[ r]} =0\qquad .\label{Eqcruc}
 \end{align}
Moreover, by expanding Eq.  [\ref{EqVw}] in  power of $\omega$,   it splits into two parts :
\begin{align}
&n[r]\frac{\big(D[r]\,C[r](\delta_2[r]-\alpha_2[r])+C^2[r]\,\beta_2[r]-D^2[r]\,\gamma_2[r]}{A[r]\,D[r]-B[r]\,C[r]}=1\label{EqUn}\qquad ,\\
&n[r]\,\frac{(C'[r]+\alpha_0[r]\,C[r]+\gamma_0[r]\,D[r])D[r]-(D'[r]+\delta_0[r]\,D[r]+\beta_0[r]\,C[r])C[r]}{A[r]\,D[r]-B[r]\,C[r]}=V[r]\qquad .\label{EqV}
\end{align}
 
 To go ahead let us assume $\gamma_2[r]\neq 0$. We obtain from the condition [\ref{aAgB}]  that :
 \begin{align}
 B[r]=-\frac{\alpha_2[r]}{\gamma_2[r]}\,A[r]\label{AdeB}  {=:\mu[r]\,A[r]}\qquad ,
 \end{align}
   {(where we have introduced $\mu[r]:=-\alpha_2[r]/ \gamma_2[r]$ to lighten the next equation). Indeed, by introducing this expression of $B[r]$ into Eqs [\ref{EqC}, \ref{EqD}] and substituting the results in Eq. [\ref{ADBC}] we obtain the main equation of the reduction process :
 \begin{align} 
  A^2[ r]=    \frac{e^{-\int  (\alpha_0[ r]+\delta_0[ r])\,dr}}  {n[r]\big(\beta_0[r]+(\alpha_0[r]+\delta_0[r])\,\mu[r]+\gamma_0[r]\,\mu^2[r]+\mu' [r]\big) }\label{EqK}\qquad .
\end{align}
}
To simplify some expressions we find useful to set :
\begin{align}
A [r]=:\frac{ \tilde A[r]}{\sqrt{n[r]}}\qquad ,\qquad B [r]=:\frac{ \tilde B[r]}{\sqrt{n[r]}}\qquad , \label{ABtAtB}
\end{align}
  {and rewrite Eq.  [\ref{AdeB}] as :
\begin{align}
 \tilde B[r]=\mu[r]\,\tilde A[r]\qquad .\label{tAdetB}
\end{align}
}
  { Inserting these redefinitions into Eqs [\ref{EqC}, \ref{EqD}] we obtain} :
\begin{align}
C[r]=
 &-\frac 12\tilde A[r]\frac{n'[r]}{n^{1/2}[r]}+n^{1/2}[r]\,\tilde C[r]\qquad , \label{CtAtC}\\
D[r]=
 &-\frac 12\tilde B[r]\frac{n'[r]}{n^{1/2}[r]}+n^{1/2}[r]\,\tilde D[r] \qquad ,\label{DtAtD}
\end{align}
  {in which we have introduced the functions $\tilde C[r]$ and $\tilde D[r]$ defined as :
\begin{align}
\tilde C[r]:=  \tilde A'[r]+\alpha_0[r]\tilde A[r]+\gamma_0[r]\tilde B[r] \qquad , \label{tCdetA}\\
\tilde D[r]:= \tilde B'[r]+\delta_0[r]\tilde B[r]+\beta_0[r]\tilde A[r] \qquad .\label{tDdetA}
\end{align}
}
All these redefinitions allow  us to rewrite Eq.  [\ref{EqUn}] as :
\begin{align}
&  n^2[r]= \frac{\tilde A[r]\,\tilde D[r]-\tilde B[r] \,\tilde C[r]}{\Big(\beta_2[r]\,\tilde C^2[r]-\gamma_2[r]\,\tilde D^2[r]-2\,\alpha_2[r]\,\tilde C[r]\,\tilde D[r]\Big)} \label{Eqn2}
\end{align}
which is easily expressed only in terms  of $K[r]:=n[r] A^2[r]=\tilde A^2[r]$ using :
\begin{align}
&\tilde B[r]\,\tilde C[r]=\mu[r]\,\tilde A[r]\,\tilde C[r]=\mu[r] \Big(\frac 12\,K'[r]+\big(\alpha_0[r]-\gamma_0[r]\,\mu[r]\big)\,K[r]\Big)\qquad ,\\
&\tilde A[r]\,\tilde D[r]= \frac {\mu[r]}{2}\,K'[r]+\Big(\beta_0[r]+\delta_0[r]\,\mu[r]+\mu'[r]\Big)\,K[r]\qquad ,
\end{align}
  {(that are immediate consequences of Eqs [\ref{tCdetA}, \ref{tDdetA}])}
and
\begin{align}
&\tilde C^2[r]=\frac1{K[r]}\Big(\tilde A[r]\,\tilde C[r]\Big)^2\qquad ,\\
&\tilde D^2[r]=\frac1{K[r]}\Big(\tilde A[r]\,\tilde D[r]\Big)^2\qquad .
\end{align}
  {The expression of the function $K[r]$ is given by  Eq.  [\ref{EqK}].}\newline
Accordingly the effective potential is completely determined by the functions appearing in the first order differential system [\ref {Eqsyst1}, \ref {Eqsyst2}].    {More precisely, the differential system Eqs[\ref{Eqsyst1}, \ref{Eqsyst2}] may be reduced to a Schr\"odinger-type system (Eqs[\ref{Sch1}, \ref{Sch2}]) under the assumptions [\ref{Eqab}, \ref{Eqcd}]   if the condition [\ref{conda2d2} is satisfied. In this case the function $n[r]$ and  the functions $A[r]$, $B[r]$, $C[r]$ and $D[r]$ defining a linear transformation between the solutions of the original differential system and the Schr\"odinger-type one are obtained via the help of a function $K[r]$ given by Eq.  [\ref{EqK}]. From this function (defined up to a multiplicative constant) we first obtain the function  $\tilde A[r]$ (as a square root of $K[r]$) and then the three functions $\tilde B[r]$,$\tilde C[r]$ and $\tilde D[r]$ thanks to Eqs [\ref{tAdetB}, \ref{tCdetA}, \ref{tDdetA}]. Then Eq.  [\ref{Eqn2}] provides (up to an arbitrary sign) the function $n[r]$. from which, using Eqs [\ref{ABtAtB}, \ref{CtAtC}, \ref{DtAtD}] we obtain the expressions of $A[r]$,$B[r]$, $C[r]$ and $D[r]$. So we have explicit expressions of all the elements  needed to compute the potential $V[r]$ via Eq.~[\ref{EqV}].}\newline
On the contrary to what is claimed in Ref. \cite{FZ}, no arbitrary function remains as the procedure goes on but only an  integration constant and a sign, both irrelevant.
 Applying this reduction process to the system Eqs [\ref{Eqdw}, \ref{Eqdk}] derived in the framework of torsion bigravity we recover the effective potential displayed in Eq.~(5.21) of Ref.\cite{VN}. Moreover Eq.~[\ref{Eqn2}] leads to :
 \begin{align}
 n^2[r]=\Big(\frac {dr}{dr^\star}\Big)^2=\Big(\frac{(r-2\,m)}r\Big)^2
 \end{align}
 in accordance with the covariance expected for a second order perturbation equation on  a Schwarzschild background.
 \section{Conclusion}
 This note has no other claim than to provide a more direct way of obtaining the effective potential driving the spherically symmetric perturbations of a Schwarzschild black hole in torsion bigravity. All the merit of its first writing and its analysis from a physical point of view remains due to the author of the work \cite{VN}.
 \bibliographystyle{ieeetr}

\end{document}